\PassOptionsToPackage{url=false, doi=false}{biblatex}
\documentclass{dsc}

\usepackage[latin1]{inputenc} 
\usepackage{amsmath}
\usepackage{amsfonts}
\usepackage{amssymb}
\usepackage{lmodern}

\usepackage{multirow}
\usepackage{xcolor}

\ifpdf \usepackage[pdftex]{graphicx} \pdfcompresslevel=9
\else \usepackage[dvips]{graphicx} \fi

\usepackage{cleveref}
\usepackage{float}
\usepackage{dblfloatfix}
\usepackage{booktabs}
\usepackage{bbding}

\bibliography{main-full_paper.bbl}

\title[Modelling individual motion sickness]{Modelling individual motion sickness accumulation in vehicles and driving simulators}
\author[V. Kotian]
       {\textbf{Varun Kotian$^\text{1}$,
                Daan M. Pool$^\text{2}$ and
                Riender Happee$^\text{1}$}}

\begin{document}
\setcounter{page}{1}
\maketitle

\begin{affiliation}
(1) Delft University of Technology, Faculty of Mechanical, Maritime and Materials Engineering, Cognitive Robotics, 2628 CD Delft, e-mail: \{v.kotian,r.happee\}@tudelft.nl\\
(2) Delft University of Technology, Faculty of Aerospace Engineering, section Control \& Simulation, 2629 HS Delft, email: d.m.pool@tudelft.nl
\end{affiliation}

\begin{abstract}
Users of automated vehicles will move away from being drivers to passengers, preferably engaged in other activities such as reading or using laptops and smartphones, which will strongly increase susceptibility to motion sickness.
Similarly, in driving simulators, the presented visual motion with scaled or even without any physical motion causes an illusion of passive motion, creating a conflict between perceived and expected motion, and eliciting motion sickness. 
Given the very large differences in sickness susceptibility between individuals, we need to consider sickness at an individual level. 
This paper combines a group-averaged sensory conflict model (as in \cite{Wada2020a}) with an individualized accumulation model (as in \cite{Oman1990, Irmak2020, Irmak2022c}) to capture individual differences in motion sickness susceptibility across various vision conditions. 
This consideration of the effect of vision is crucial in driving simulators where there is a strong contribution of visual cues.
The model framework can be used to develop personalized models for users of automated vehicles and improve the design of new motion cueing algorithms for simulators.
The feasibility and accuracy of this model framework are verified using two existing datasets with sickening conditions in 1) an experimental vehicle with and without outside vision \citep{Irmak2020}, and 2) comparing vehicle experiments with corresponding driving simulator experiments \citep{Talsma2023ValidationResearch}. Both datasets involve passive motion, representative of being driven by an automated vehicle. The model is able to fit an individual's motion sickness responses using only 2 parameters (gain $K_1$ and time constant $T_1$), as opposed to the 5 parameters in the original model. This ensures unique parameters for each individual.
Better fits, on average by a factor of 1.7 (for Accum\_2 model), of an individual's motion sickness levels, are achieved as compared to using only the group-averaged model (Accum\_0 model).
Furthermore, this model framework demonstrates robustness by accurately modeling various datasets with distinct motion and vision conditions.
Thus, we find that models predicting group-averaged sickness incidence cannot be used to predict sickness at an individual level. On the other hand, the proposed combined model approach predicts individual motion sickness levels and thus can be used to control sickness. 


\end{abstract}

\begin{keywords}
Motion Sickness, Simulator Sickness, Modeling, Driving Simulators, Automated Vehicles.
\end{keywords}

\section*{Introduction}
Automated vehicles and driving simulators are very different technologies. However, they both share two common facts. The first is that they have become very popular in recent years, a trend that is expected to continue in the future. Secondly, they both share a common issue in \textit{motion sickness}.
Users of automated vehicles will move away from being drivers to passengers, preferably engaged in other activities such as reading or using laptops and smartphones, which strongly increases susceptibility to motion sickness.
Similarly, in driving simulators, the apparent visual motion combined with scaled or without any physical motion causes an illusion of passive motion, creating a conflict between perceived and expected motion, and eliciting motion sickness \citep{Bos2020MotionSickness}. Though these two cases seem different, the inherent mechanism that causes motion sickness in both, i.e., \textit{sensory conflict}, is the same. 

The mechanisms behind the development and evolution of motion sickness have been studied extensively, relying heavily on models that predict sensory conflicts based on inputs from the vestibular and visual sensory systems \citep{Bos1998b, Wada2020a, Liu2022, Irmak2023ValidatingPrediction, Kotian2023TheSickness}. These models are known as `conflict generation' models. However, these models predict group-averaged sickness responses (conflicts), using Motion Sickness Incidence (MSI), and thus cannot be reliably used for predicting and controlling sickness, both in real vehicles and simulators, at an individual level.

\begin{figure*}[ht]
    \centering
    \includegraphics[width=\textwidth]{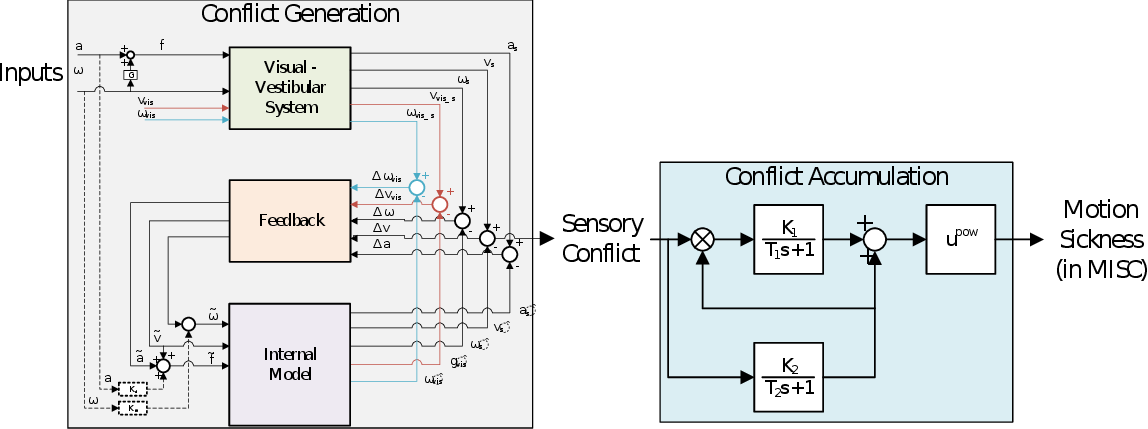}
    \caption{Framework for combined `conflict generation' \citep{Wada2020a, Liu2022} and `conflict accumulation' \citep{Oman1990, Irmak2022c} models to predict individual motion sickness, i.e., MIsery SCale \citep{Reuten2020}}
    \label{fig:combined_model}
\end{figure*}

`Conflict generation' models, such as the SVC model by \citep{Wada2020a}, calculate MSI using a conflict accumulator that integrates the generated conflicts in time. However, this is a single low-pass integrator and hence is not able to model the fast dynamics of motion sickness such as recovery.
One solution is to make use of the same `conflict generation' model and replace the accumulation integrator with a much more complex integrator model (as in \cite{Oman1990}), which will be referred to as the `conflict accumulation' model in this paper (see \cref{fig:combined_model}). 
These `conflict accumulation' models have been found very useful in motion sickness predictions for one degree of freedom motion \citep{Irmak2020, Irmak2022c}.
These will be personalized to each individual with a unique set of parameters. 
Further, a complex `conflict accumulation' model will be able to capture phenomena like recovery from motion sickness and hypersensitivity \citep{Irmak2020, Irmak2022c}.
The personalization of parameters has already been shown to improve modeling accuracy by \citep{Irmak2020} where it increased the accuracy by a factor of 2 as compared to group-averaged modeling.
Our work extends this to 6 degrees of freedom (DOF). This model has the additional benefit of allowing visual inputs. This is important when predicting motion sickness in simulators, i.e. simulator sickness, where there is a strong influence of vision. 
Additionally, individual motion sickness metrics, such as those obtained using the MISC scale \citep{DeWinkel2022}, instead of group-averaged metrics like Motion Sickness Incidence (MSI) (as in \cite{Wada2020a, Kotian2023TheSickness}), allow individualized fitting of the model resulting in better prediction in future motion paradigms for the individual. 
Hence, in this paper, we aim to combine the `conflict generation' model with a `conflict accumulation' model to improve the ability to capture the differences in individual susceptibility. This model framework is visualized in \cref{fig:combined_model}.

\section*{Research Question}
The goal of the paper is to verify the feasibility of the combined motion sickness model approach, i.e., combining `conflict generation' and `conflict accumulation' models, to capture individual differences in motion sickness susceptibility across various vision conditions. 
To do this we have two further sub-goals:
\begin{itemize}
    \item To verify the accuracy of the combined conflict generation/accumulation modeling approach
    \item To find a balance between model accuracy and the number of estimated accumulation model parameters 
\end{itemize}
The second goal ensures a practical model with estimated parameter sets unique to each individual.
We use two existing datasets with sickening conditions in 1) an experimental vehicle with and without outside vision \citep{Irmak2020}, and 2) comparing vehicle experiments with corresponding driving simulator experiments \citep{Talsma2023ValidationResearch}. Both datasets involve passive motion, representative of being driven by an automated vehicle.
    
By achieving this goal, this model framework can be used to develop personalized models for users of automated vehicles, which the automated vehicle can use to determine the comfort levels of the user and subsequently best adapt its driving style, i.e., by limiting the acceleration and rotations of the car. This model framework can also improve and accelerate the design of new optimization-based motion cueing algorithms that aim to minimize sickening motions in simulators like in \cite{Hogerbrug2020SimulatorEnvironments, Baumann2021DrivingVehicles}. Here, the tilting to replicate specific forces and/or visual cueing delays will be optimized, which will result in fewer dropouts of participants due to motion sickness.

\section*{Methodology}

\begin{table*}[h]
\centering

\caption{Experimental datasets used in this study}
\label{tab:exp_data}

\resizebox{0.9\linewidth}{!}{%
\begin{tabular}{@{}l|l|l|c@{}}
\toprule
Datasets              & Details               & Reference   & No. of Participants\\ \midrule
\begin{tabular}[c]{@{}l@{}}Slalom Drive\\ Dataset\end{tabular} &
  \begin{tabular}[c]{@{}l@{}}Slalom with Internal and External vision\\ Motion sickness responses for hypersensitivity\end{tabular} &
  \cite{Irmak2020} & 16\\
           &         &          \\
\begin{tabular}[c]{@{}l@{}}Car and Simulator\\ Dataset\end{tabular} &
  \begin{tabular}[c]{@{}l@{}}Naturalistic drive in vehicle and moving base simulator\\ Only External vision\end{tabular} &
  \cite{Talsma2023ValidationResearch} & 24\\
  \bottomrule
\end{tabular}
}
\end{table*}

\subsection*{Experimental Datasets}
\begin{figure}[h]
    \centering
    \includegraphics[width=.75\linewidth]{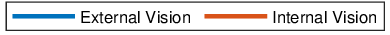}
    \includegraphics[width=.95\linewidth,trim={0.75cm 0 0.75cm 0},clip]{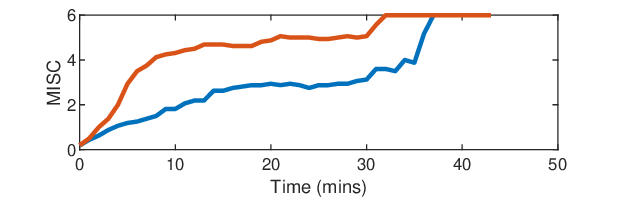}
    \caption{Slalom Drive dataset \citep{Irmak2020} group-averaged MISC levels versus time in the conditions of external (blue) and internal vision (red)}
    \label{fig:slalom_mean_misc}

    \includegraphics[width=.5\linewidth]{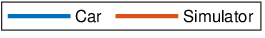}
    \includegraphics[width=.95\linewidth,trim={0.75cm 0 0.75cm 0},clip]{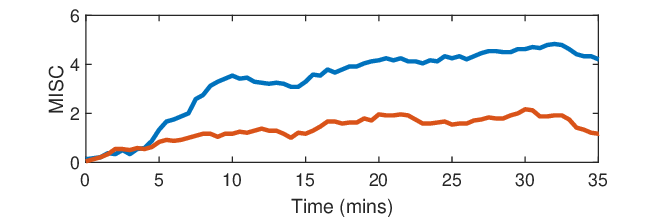}
    \caption{Car and Simulator dataset \citep{Talsma2023ValidationResearch} group-averaged MISC levels versus time in the car (blue) and simulator (red)}
    \label{fig:talsma_data}
\end{figure}

For the analysis in this paper, we make use of an existing real-world `Slalom Drive' dataset (16 participants) with varying vision conditions, i.e., with and without an outside view, where mean MISC levels at the end of motion exposure were 5.3 (severe symptoms) with an outside view and 3.3 (some symptoms) without an outside view (Experiment 1 in \citep{Irmak2020}), see \cref{tab:exp_data} and \cref{fig:slalom_mean_misc}. This experiment compared the motion sickness development with and without an outside view from the car. This dataset is ideal for proving that our new model framework can simulate various vision conditions in actual vehicles.
The model framework is also validated using the `Car and Simulator' dataset (24 participants) by \cite{Talsma2023ValidationResearch}. This experiment contains motion sickness responses from real-world driving and its (matched) simulation on a moving-base driving simulator \citep{Talsma2023ValidationResearch}. For this experiment, the mean MISC levels at the end of motion exposure were around 5.5 (severe symptoms) in the car and 1.5 (slight discomfort or vague symptoms) in the simulator (see \cref{tab:exp_data} and \cref{fig:talsma_data}).
Using the Car and Simulator dataset, we demonstrate the capability of the model framework to generalize well for different types of passive motion, whether it originates from a car or a simulator.
Furthermore, both datasets are used to study the number of parameters needed to accurately simulate the motion sickness development in different participants.

Both datasets capture sickness responses to passive motion, representative of being driven by an automated vehicle as well as driving simulators. In these datasets individual motion sickness levels (using MISC) were reported as a function of time with motion stimuli varying in time, showing major individual differences in sickness susceptibility, making this data suitable to model sickness accumulation. 

\subsection*{Inputs and Outputs}
Inertial inputs (on the left side of \cref{fig:combined_model}), such as acceleration and angular velocity, and vision inputs (on the left side of \cref{fig:combined_model}), such as visual verticality (orientation) and visual rotation (rotational velocity), will be input into this model framework. The outputs (MISC) will be compared to the ground truth from the experiments and the model parameters (time constants and gains, explained in detail in the next section) in the `conflict accumulation' model will be tuned to have the best fit to individual responses. More details about each model are given in the following subsections. We applied 6D motion (3D translation, and 3D rotation), using the recorded seat and platform motion for the Car and Simulator dataset, and recorded head motion for the Slalom Drive dataset.

The human eye estimates motion through vision by measuring the rotation of visual cues between the current and previous states, which is known as optic flow. Therefore, we made the assumption that the visually perceived rotations would be equivalent to head (or vehicle, as in the case of the Car and Simulator dataset) rotations when observing the external environment. For internal vision, we select a zero visual input assuming no head motion relative to the vehicle.

\subsection*{Selection of models}

As previously discussed, we combine two models in our proposed model framework. The first component of the model is the `conflict generation' model (see the left part in \cref{fig:combined_model}), which models the integration of sensory inputs and generates a one-dimensional sensory conflict signal. 
This model should be reliable in forecasting conflict signals consistently, especially while driving around in a vehicle. 
Additionally, the model must be capable of incorporating visual inputs so that the effect of different vision conditions can be studied.
To select an appropriate model, we refer to a study by \cite{Kotian2023TheSickness} where various models and implementations of vision inputs were compared.
Based on their findings, we conclude that the 6DOF Subjective Vertical Conflict (SVC) model with Visual Rotational Velocity input \citep{Wada2020a} is ideal for our purpose of predicting motion sickness.
This model has been shown to accurately replicate the frequency and amplitude dynamics of motion sickness as reported in numerous literature studies. It is shown that the Visual Vertical (red pathways in \cref{fig:combined_model}) input as implemented in the SVC model was not effective in predicting sickness and hence will not be used in the current study (and set the corresponding gains to zero).

The second part of the model is the `conflict accumulation' model. As the name suggests, this model accumulates (integrates) the sensory conflicts generated by the first part of the model framework to estimate the level of motion sickness experienced by an individual as it builds up over time.
We adopt the more advanced Oman model \citep{Oman1990} for this. The Oman model is a five-parameter non-linear model that integrates the conflict (see the right part in \cref{fig:combined_model}), with fast and slow pathways combined with leakage.
The fast path has a very low time constant ($T_1$) and models the direct response to a sickening stimulus. The slow path has a high time constant ($T_2$) and captures the long-term effects of sickening stimuli, such as recovery and hypersensitivity. 
This is also relevant in simulators where a sudden increase in motion incongruence can make participants hypersensitive. This has been observed by \cite{Cleij2018ContinuousSimulation, Kolff2022MotionCueing}, where it was shown that motion tends to be momentarily bad, but not continuously.
Both pathways have a gain ($K_1$ and $K_2$) to control the contribution of each path. Additionally, there is a power law ($pow$) at the model's output to account for nonlinear scaling effects.

 In this paper, we use the combined model to predict sickness as rated with the MIsery SCale (MISC) \citep{Reuten2020, DeWinkel2022}.
The MISC is an 11-point symptom-based scale (0-10) that is used to query motion sickness in various studies (such as in \cite{Irmak2020, Irmak2022c, Talsma2023ValidationResearch}).

\subsection*{Parameter reduction in the Accumulation model}
In addition to studying the accuracy of the Accumulation model as it is (with 5 individual parameters), we also tried to reduce the number of parameters needed while ensuring goodness of fit. This will prevent over-fitting and result in unique solutions. 
To do this we use empirical relations and median values of parameters as observed in previous studies and current simulations with various numbers of parameters. 
\begin{itemize}
    \item $K_2 = 5 K_1$; from \cite{Oman1990}
    \item $T_2 = 7 T_1$; from \cite{Irmak2020}
    \item $pow = 0.4$; from \cite{Irmak2022c}
    \item $T_1 = 60$ s; from \cite{Oman1990}
    \item $K_1$ best fit within datasets (median values from individual estimates)
    \begin{itemize}
        \item $K_1 = 2$; for Slalom Drive dataset
        \item $K_1 = 18$; for Car and Simulator dataset 
    \end{itemize}    
\end{itemize}
We use combinations of these assumptions to test if the number of parameters can be reduced.
\Cref{tab:no_of_params} summarizes the different cases of the accumulation (Oman) model with the parameters that are estimated (marked with a \Checkmark) and the parameters that are held constant or as an empirical relation to the estimated parameters. $K_1$ and $K_2$ are the gains, $T_1$ and $T_2$ are the time constants and $pow$ is the power term in the accumulation model in \cref{fig:combined_model}. Accum\_5 is the original version of the accumulation model from \cite{Oman1990}. Accum\_0 is the model with group-averaged parameters. The $K_1$ is different for each dataset as the motion captured in each dataset is different. In the `Slalom Drive' head motion was recorded, however, in the `Car and Simulator' only vehicle motion was recorded. Hence, there is a difference in the magnitude of rotations which explains the difference in estimated gains ($K_1$).

\begin{table}[h]
\centering
\caption{Parameter reduction. Simplified versions of the `conflict accumulation' model with a different number of parameters. Check marks indicate that the parameter is estimated for each individual.}
\label{tab:no_of_params}
\footnotesize
\begin{tabular}{@{}lcccccc@{}}
\toprule
Label    & \# parameters  & $K_1$ & $K_2$   & $T_1$ & $T_2$   & $pow$ \\ \midrule
Accum\_5  & 5                 &  \Checkmark  &   \Checkmark   &  \Checkmark  &   \Checkmark   &   \Checkmark  \\
Accum\_4a & 4                 &  \Checkmark  & $5 K_1$ &  \Checkmark  &   \Checkmark   &  \Checkmark   \\
Accum\_4b & 4                 &  \Checkmark  &   \Checkmark   &  \Checkmark  & $7 T_1$ &  \Checkmark   \\
Accum\_3  & 3                 &  \Checkmark  & $5 K_1$ &  \Checkmark  & $7 T_1$ &  \Checkmark   \\
Accum\_2  & 2                 &   \Checkmark & $5 K_1$ &  \Checkmark  & $7 T_1$ & 0.4 \\ 
Accum\_1a & 1                 &  2 or 18  & $5 K_1$ &  60  &   \Checkmark   &  0.4   \\
Accum\_1b & 1                 &  2 or 18 &   \Checkmark   &  60  & $7 T_1$ &  0.4   \\
Accum\_0 & 0                 &  2 or 18 &  $5 K_1$  &  60  & $7 T_1$ &  0.4   \\
\bottomrule
\end{tabular}
\end{table}

\subsection*{Parameter estimation in the accumulation model}

To allow the `conflict accumulation' model to accurately capture the individual differences in motion sickness susceptibility, the parameters need to be set for each individual. Further, to model both conditions together, only one set of parameters will be estimated for both conditions for each individual. 
To fit these parameters a constrained optimization problem is formed. 
This estimation is carried out in \textit{MATLAB} with the \textit{fmincon} solver using the \textit{sqp} algorithm. In addition to this, \textit{multistart} was used to simultaneously find 16 local minima and then find the lowest out of them. This way it is ensured that we find  the global minimum. 
The Root-Mean-Squared Error (RMSE) between the actual and predicted MISC responses, as a function of the parameter vector $x$, is chosen as the cost function for the estimation. 
As both conditions are fitted together to give one set of parameters, the cost function is the sum of RMSE for both conditions. 
The optimization problem along with the constraints are shown in \cref{eq:cost_func}. 
\begin{equation}
\begin{array}{cc}
\underset{x}{\text{minimize}} & RMSE_{C_1}(\boldsymbol{x}) + RMSE_{C_2}(\boldsymbol{x}),  \\ 
\text{where}, & x = (K_1, K_2, T_1, T_2, pow)^T
\end{array}
\label{eq:cost_func}
\end{equation}

\section*{Results}

\begin{figure*}[h]
    \centering    
    \hspace{0.07\linewidth}\includegraphics[width=0.7\linewidth]{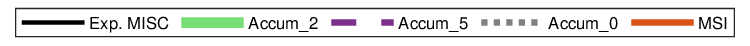}
    \includegraphics[width=0.9\linewidth]{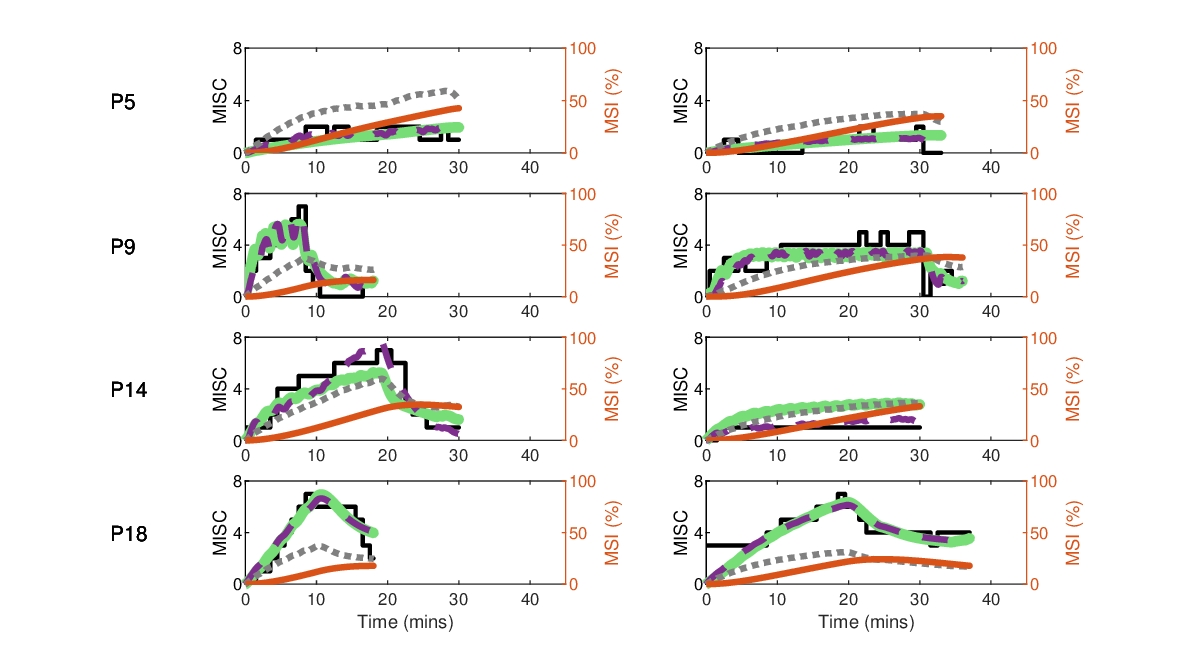}
    \caption{Slalom Drive vehicle tests. Motion sickness responses (MISC) in experiments by \cite{Irmak2020} in black, fitted Accum\_2 model predictions (MISC) in green, fitted Accum\_5 model predictions (MISC) in dashed violet, fitted Accum\_0 model predictions (MISC) in dotted grey, and MSI predictions from Hill function in orange for 4 participants (participant label shown on the left) for the conditions of external (right column) and internal (left column) vision.}
    \label{fig:slalom_fits}
\end{figure*}

In this paper, we propose a new model framework that combines the `conflict generation' model with an advanced `conflict accumulation' model to improve the ability to capture differences in individual susceptibility.
Using the methods described in the previous section, simulations were run and the results are presented below.
These results are presented with the aim to justify our goal, i.e, to verify the feasibility and the accuracy of the new model framework.
It is hypothesized that using a `conflict accumulation' model that uses individual parameters will be more accurate than a `conflict accumulation' model that uses group-averaged parameters, as also reported by \citep{Irmak2020}. 
First, results are shown for the Slalom Drive dataset by \cite{Irmak2020} to show the performance of the model with varying vision conditions. This is followed by results for the Car and Simulator dataset by \cite{Talsma2023ValidationResearch} where the adaptability of the model to real-world driving and driving simulators is shown.
Fits of the model framework to the actual MISC responses with Motion Sickness Incidence (MSI) predictions overlayed are shown first followed by their RMSE with the actual MISC responses.

\Cref{fig:slalom_fits} show the comparison of the various models with experimental recorded motion sickness responses (MISC).
It is evident that our approach of estimating parameters for each individual (in particular for Accum\_2 and Accum\_5 models) offers improved accuracy in predicting MISC responses compared to the use of group-averaged parameters (Accum\_0). The average RMSE reduces from 1.13 and 0.74, for Accum\_2 and Accum\_5 models, to 1.94, for the Accum\_0 model. This proves that using parameters estimated for each individual (Accum\_2 and Accum\_5) is, on average, 1.7 times better than using group-averaged parameters, as in the Accum\_0 model.

Another important observation is that all the accumulation models capture the recovery from motion sickness. This recovery occurs when the sickening stimuli are stopped and the participant is allowed to rest. This is more evident in participants 9 and 14 (second and third row). The Hill function predicting MSI (in orange) is not able to capture this reduction of motion sickness. MSI reports the percentage of people who might get motion sick but does not give information about any certain individual. 

\begin{figure}[h]
    \centering
    \includegraphics[width=0.8\linewidth]{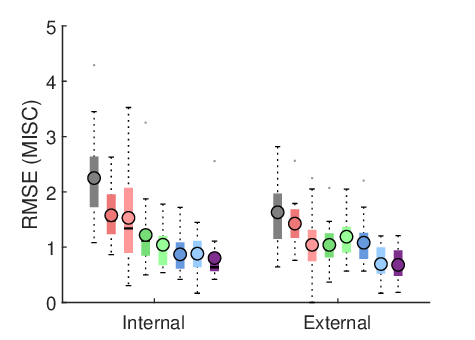}
    \vskip-13em
    \hspace{4em}
    \includegraphics[width=0.55\linewidth]{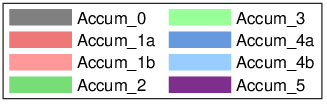}
    \vskip8em
    \caption{Root Mean Squared Error (RMSE) between the predicted MISC and the actual MISC for the Slalom Drive dataset. Shown are the mean (circle), median (horizontal solid line), and interquartile range (colored rectangle)}
    \label{fig:rmse_oman_slalom}
\end{figure}

\begin{figure*}[b]
    \stepcounter{figure}
    \centering    
    \hspace{0.07\linewidth}\includegraphics[width=0.7\linewidth]{fits_legend.eps}
    \includegraphics[width=0.9\linewidth]{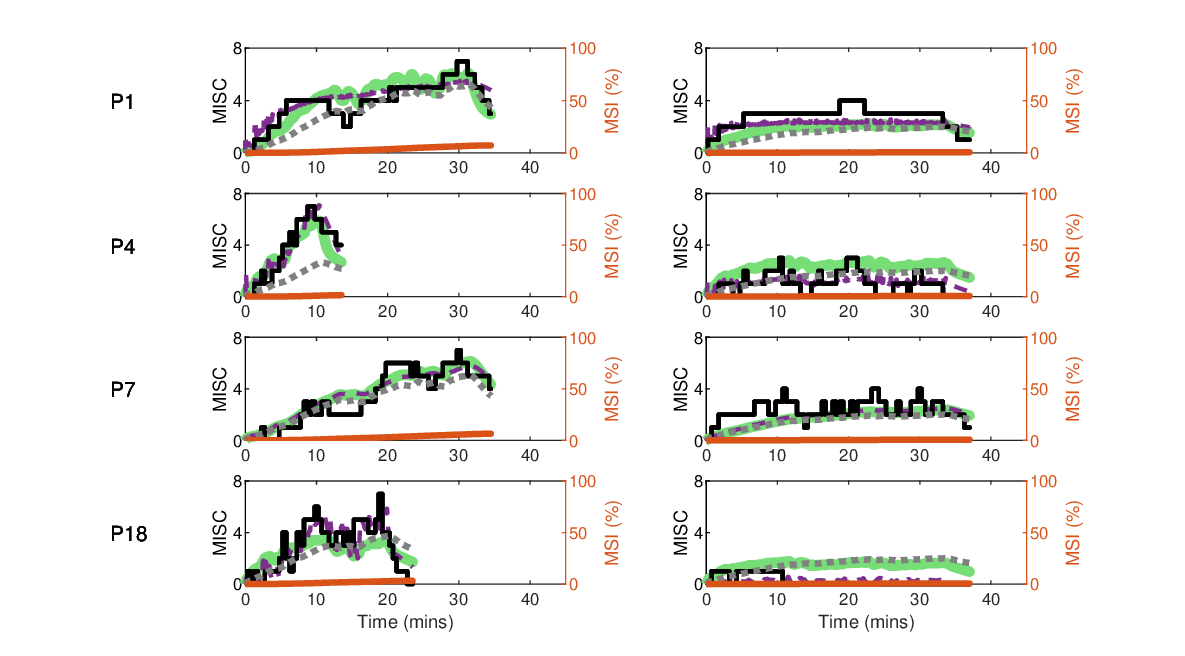}
    \caption{Car and Simulator tests. Motion sickness responses (MISC) in experiments by \cite{Talsma2023ValidationResearch} in black, fitted Accum\_2 model predictions (MISC) in green, fitted Accum\_2 model predictions (MISC) in dashed violet, fitted Accum\_0 model predictions (MISC) in dotted grey, and MSI predictions from Hill function in orange for 4 participants (participant label shown on the left) for the case in the car (left column) and the simulator (right column) vision.}
    \label{fig:talsma_datset_fits}
\end{figure*}

Additionally, we also evaluated the need for each of the parameters. We did this by reducing the number of parameters by using relations found during the previous simulations and also what was observed and reported in various studies. We used 5, 4, 3, 2, 1, and 0 parameters to evaluate the need for the parameters, see \cref{tab:no_of_params}.
From \cref{fig:rmse_oman_slalom} it is clear that reducing the number of parameters below 2 leads to a 36\% increase in RMSE (from 1.2 to 1.6 in internal vision case and from 1.0 to 1.4 in external vision case when comparing Accum\_2 to Accum\_1a model). 
Hence, any model with 2 or more parameters is sufficiently accurate to capture the motion sickness development. We also compared the individual models with a group-averaged version of the accumulation model (Accum\_0), where the parameters are the same for all participants in the dataset. 
It is observed that Accum\_0 has on average 1.7 times more RMSE (from 2.25 to 1.22 in the internal and from 1.63 to 1.04 in the external vision case) as compared to the Accum\_2 model. Thus, parameters estimated for each individual are more accurate than group-averaged parameters. 

\Cref{fig:PP_slalom} shows the distribution of the two estimated parameters ($K_1$ and $T_1$) in the Accum\_2 model for the Car and Simulator dataset. The black line shows the parameters for the Accum\_0 model as a reference.
It can be observed in that there is a wide range of parameter combinations (median: 2 and 83.5, standard deviation: 3.8 and 190.5, for $K_1$ and $T_1$, respectively) describing individuals that show variations in motion sickness susceptibility. These parameter sets can be classified into three groups of motion sickness susceptibility: high, medium, and low.
High susceptibility are those with large $K_1$ and small $T_1$. Low susceptibility have the opposite, small $K_1$ and large $T_1$. Lastly, medium susceptibility have small $K_1$ and small $T_1$. With this knowledge, representative parameters can be generated to test motion profiles on different motion sickness susceptibilities. Additionally, percentiles can be defined to see which percentile of subjects do or do not get motion sick.

\begin{figure}[h]
    \addtocounter{figure}{-2}
    \centering
    \includegraphics[width=0.5\linewidth]{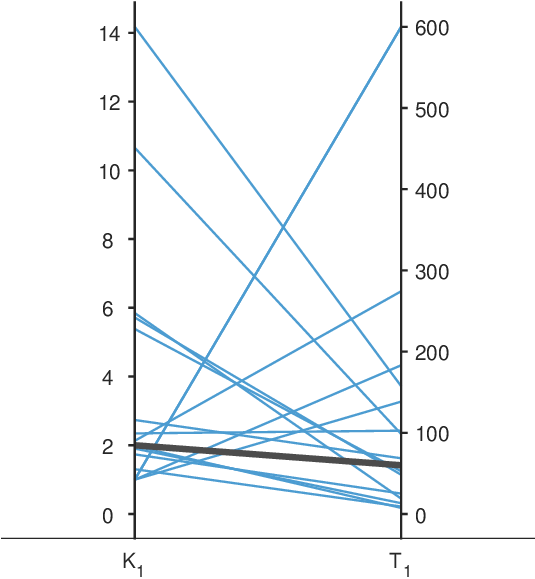}
    \caption{Slalom drive dataset parameter distribution (estimated gain ($K_1$) and time constant ($T_1$)) for the Accum\_2 model (blue) and Accum\_0 model (black)}
    \label{fig:PP_slalom}
    \addtocounter{figure}{1}
\end{figure}

Furthermore, we validated these models on the car and simulator dataset by \cite{Talsma2023ValidationResearch}. This dataset had 24 participants, each experiencing motion with external vision in a real-world car and in a simulator. \Cref{fig:talsma_datset_fits} shows the experimental recorded motion sickness responses (MISC) in black, MSI predictions in orange, and fitted model predictions in green, violet, and grey  for 4 out of the 24 participants for both cases. 
It can be seen that even with 2 parameters, the fits of the Accum\_2 model are very close to the actual MISC responses in both the car (RMSE mean: 1.03, std:0.544) and simulator (RMSE mean:1.06, std: 0.605). The MSI prediction values are very low. This is due to the low levels of conflict generated in these experiments as compared to the slalom drive by \citep{Irmak2020}. Another reason may be that the participants sampled in \cite{Talsma2023ValidationResearch} may be highly susceptible to motion sickness. The `conflict Accumulation' model is able to account for these combinations of low levels of conflict and high susceptibility of the participants adequately.

\begin{figure}[h]
    \centering
    \includegraphics[width=0.8\linewidth]{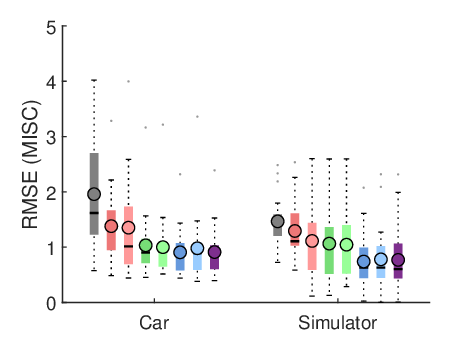}
    \vskip-13em
    \hspace{4em}
    \includegraphics[width=0.55\linewidth]{RMSE_boxplot_legend.eps}
    \vskip8em
    \caption{Root Mean Squared Error (RMSE) between the predicted MISC from the models and the actual MISC for the Car and Simulator Dataset. Also shown are the mean (circled), median (horizontal solid line), and interquartile range (in colored rectangle)}
    \label{fig:rmse_oman_talsma}
\end{figure}

Lastly, to demonstrate that the two parameters are the most optimal number of parameters even for this dataset, this was tested with other versions of the model with varying numbers of parameters. As shown in \Cref{fig:rmse_oman_talsma}, there is a 25\% increase in RMSE values (from 1.03 to 1.38 in the case of motion sickness in the car and from 1.06 to 1.29 in the simulator) when comparing Accum\_2 to Accum\_1a.
When comparing the individual models with a group-averaged version of the accumulation model, it is observed that the group-averaged model, Accum\_0, has 1.64 times more RMSE (from 1.96 to 1.03 in the internal and from 1.47 to 1.06 in the external vision case) as compared to the Accum\_2 model. This increase is close to what was seen in the Slalom Drive dataset where a 1.7 times increase was observed. Thus, parameters estimated for each individual are more accurate than group-averaged parameters. 

\begin{figure}[h]
    \centering
    \includegraphics[width=0.5\linewidth]{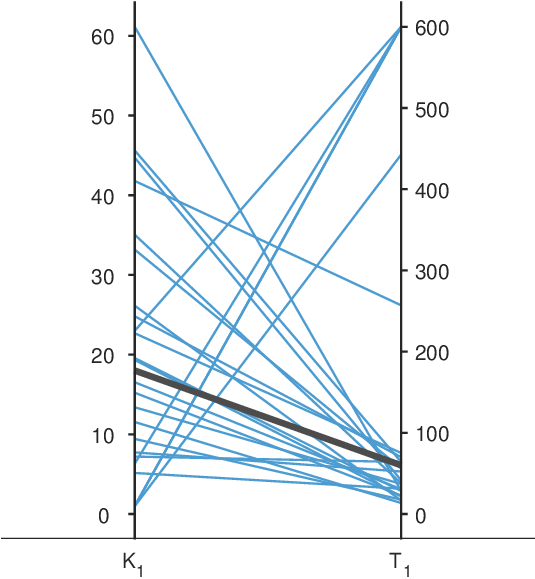}
    \caption{Car and Simulator dataset parameter distribution (estimated gain ($K_1$) and time constant ($T_1$)) for the Accum\_2 model (blue) and Accum\_0 model (black)}
    \label{fig:PP_talsma}
\end{figure}

\Cref{fig:PP_talsma} shows the distribution of the two estimated parameters ($K_1$ and $T_1$) in the Accum\_0 and Accum\_2 model for the Car and Simulator dataset. The observations are equivalent to those reported from \cref{fig:PP_slalom}. 
Individual variations in motion sickness susceptibility are shown by a wide range of parameter combinations (median: 18 and 47, standard deviation: 16.1 and 221.8, for $K_1$ and $T_1$, respectively).
However, there is a difference in the gain value ($K_1$), due to the different input data used for both datasets. In the Slalom Drive dataset, head motion is used while in the Car and Simulator dataset vehicle motion is used as the head motion was not available. Ideally, we would like to use head motion as those motions are perceived by the vestibular and visual systems.  

It is clear from the results in \cref{fig:slalom_fits,fig:talsma_datset_fits} that this model framework, is able to capture two different conditions with a single set of 2 parameters. This is applicable for various vision conditions, as well as different motions from real cars and simulators.

\section*{Discussion}
This paper introduces a novel model framework to predict an individual's motion sickness level in vehicles and simulators. This model framework combines a group-average `conflict generation' model with an individual `conflict accumulation' model to better capture individual susceptibility differences. This is important to understand and capture differences in motion sickness susceptibility.
Furthermore, by using a `conflict generation' model that includes visual inputs, various vision conditions (such as external, internal, and only vision) can be simulated as well. This is crucial in simulators where motion sickness occurs due to a strong influence of visual cues.
We hypothesized that using a `conflict accumulation' model that uses individualized parameters will result in greater accuracy compared to a model that uses group-averaged parameters. Hence, in this paper, we assessed the feasibility and accuracy of this new model approach for motion sickness predictions. 
    
It is clear from the results (see \cref{fig:slalom_fits,fig:talsma_datset_fits}) that the Oman model as a `conflict accumulation' model with individualized parameters is able to better model the motion sickness responses of individuals as compared to using the group-averaged models such as the Accum\_0.
In addition to this, it also captures the recovery phase of the experiment (see \cref{fig:slalom_fits,fig:talsma_datset_fits}) and, theoretically, the hypersensitivity in the following second motion exposure. This recovery phase takes a few minutes and is not captured by the Hill function, which predicts MSI, due to its time constant of 12 min.
These results are in line with the work by \cite{Irmak2020} where individualized fits with the Oman model reduced the prediction error by a factor of 2 in a slalom drive with a frequency of 0.2 Hz and lateral accelerations with a peak amplitude of 0.4 g with eyes closed. 
Our work extends this, by using a six degrees of freedom `conflict generation' model to generate the conflict, in realistic driving conditions with varying vision conditions. 
This way a single set of parameters, estimated using data from all required conditions, can characterize an individual across various motion and vision conditions.

We also looked into reducing the number of parameters in the model. 
We found that using the relations and values mentioned in previous studies (such as in \cite{Oman1990, Irmak2020, Irmak2022c}), the number of individual parameters can be limited to 2. 
This way we can be sure that the parameter set estimated is unique for each individual and is not an over-fit on the available sparse (in conditions) dataset.
Another by-product is that the computation time required for the estimation of parameters is significantly reduced (by a factor of 4, from 48 to 11 seconds for 40 min of simulation). However, even with 5 parameters the model can be used for real-time applications.

Using these models, the motion profiles in simulator experiments can be tuned to reduce dropout of participants in simulator experiments due to motion sickness.
Not only is the model useful in designing experiments but also can be employed in motion cueing algorithms in driving simulators, as well as in automated vehicles. Using a model-based control method, this model framework can be included in the plant model to forecast motion sickness levels, as in \cite{Jain2023OptimalAssessment}. This way, motion sickness levels can be controlled by taking into account each individual's susceptibility. Vehicle motion in automated vehicles and platform motion and tilt coordination in simulators can be optimized to minimize their effect in eliciting motion sickness.
Thus, this model is feasible to use both post-experiment and in real-time during the experiment.



We now tuned individual parameters of the accumulation model while keeping the many parameters of the sensory conflict model constant. 
A main limitation of the model is that it fails to accurately fit the subject's responses who get highly motion sick during internal vision and do not get sick during the external vision case (for example P14 in \cref{fig:slalom_fits}). This sharp shift in motion sickness dynamics cannot be captured by our model framework. To do this, at least one additional parameter needs to be estimated, which could be the vision gain in the `conflict generation' model or other perception parameters. 
People not only differ in motion sickness susceptibility, but also in their perception of motion.
This has been shown in \cite{Irmak2023ValidatingPrediction}, where a correlation coefficient of 0.74 was observed between individuals' overall sickness sensitivity and their subjective vertical time constant.
Thus each human will have a different contribution of visual and vestibular signals for their state estimation. By adjusting the parameters in the `conflict generation' model, the contribution of vision to the estimates can be tuned. 

Another limitation is that the model gain, $K_1$, needs to be adapted for the two datasets, and presumably depending on the location of the IMU (head or seat). This can be seen in the difference in the range of estimated gain ($K_1$) values in the two datasets (see \cref{fig:PP_slalom,fig:PP_talsma}). 
Also, the Hill function accumulation model predicted a reasonable MSI magnitude in the Slalom Drive but highly underestimated sickness in the Car and Simulator dataset.
Ideally, we would like to always use recorded head motion, which is more representative of the motion experienced by the vestibular system.  If this is not available we could use biomechanical human/seat models or linear transfer functions to convert 6D vehicle motion to 6D head motion. This way it can be ensured that the estimated parameters can be used for any case.


\section*{Conclusion}
A new model framework was developed to fit and predict individual motion sickness levels in vehicles and simulators. This model framework contained two parts: `conflict generation' and `conflict accumulation'. 
The model utilizes acceleration and angular rotational data as inputs and adjusts the parameters of the `conflict accumulation' model to fit each individual's motion sickness level (reported using MIsery SCale (MISC)) response.
This framework is able to fit individual responses with just two estimated parameters (a gain $K_1$ and time constant $T_1$) with an average RMSE of 1.1 MISC.
By combing the two models, better fits, on average by a factor of 1.7 (for Accum\_2 model), of an individual's motion sickness levels are achieved as compared to using only the group-averaged models.
We fit both conditions (External and Internal vision conditions for the Slalom Drive dataset and driving in a car and a simulator in the Car and simulator dataset) together to estimate one set of parameters.
The model is able to simulate both conditions using a single set of estimated parameters for each participant. 
This framework has the potential to show the impact of vision on an individual level, offering a more personalized approach compared to previous models. This is important especially in driving simulators, as the effect of delays in visual cues, can be quantified with this model.
This model framework also shows robustness by accurately modeling different datasets with completely different motion and vision conditions.

These individual models can be used in further studies for predicting motion sickness levels of the same individual in a different motion scenario, thus reducing the dependence on extensive experiments on humans. 
Finally, this model framework can also aid in faster testing of new motion cueing algorithms for driving simulators due to less sickening driving simulator experiments and may reduce the dropout of participants in driving simulator evaluations.

\section*{Acknowledgements}
We thank Tugrul Irmak (Delft University of Technology, The Netherlands) and Tessa Talsma (Skoda Auto, Czechia) for sharing the experimental data. The contribution of Varun Kotian was financially supported by Toyota Motor Europe.



\printbibliography


\end{document}